\documentclass[12pt,preprint]{aastex}

\RequirePackage{graphicx}

\newcommand{\etal}{{\sl et~al.~}}

\slugcomment{To appear in Astrophysical Journal}

\shorttitle{The Mass of the Candidate Exoplanet Companion to HD\,136118 from Hubble Space Telescope Astrometry and High-Precision Radial Velocities}
\shortauthors{Martioli et al.}

\begin{document}

\title{The Mass of the Candidate Exoplanet Companion to HD\,136118 from Hubble Space Telescope Astrometry and High-Precision Radial Velocities \footnote{Based on observations made with the NASA/ESA \textit{Hubble Space Telescope}, obtained at the Space Telescope Science Institute, which is operated by the Association of Universities for Research in Astronomy, Inc., under NASA contract NAS5-26555. Based on observations obtained with the Hobby-Eberly Telescope, which is a joint project of the University of Texas at Austin, the Pennsylvania State University, Stanford University, Ludwig-Maximilians-Universitt Mnchen, and Georg-August-Universit\"{a}t G\"{o}ttingen.}}

\author{Eder Martioli\altaffilmark{1}, Barbara E. McArthur\altaffilmark{2}, G. Fritz Benedict\altaffilmark{2}, Jacob L. Bean\altaffilmark{3}, Thomas E. Harrison\altaffilmark{4,5} and Amber Armstrong\altaffilmark{1}}
\email{edermartioli@gmail.com}

\altaffiltext{1}{Divis\~ao de Astrof\'isica, Instituto Nacional de Pesquisas Espaciais, S. J. dos Campos, SP, Brazil.}
\altaffiltext{2}{Department of Astronomy, University of Texas at Austin, TX 78712.}
\altaffiltext{3}{Institut f\"{u}r Astrophysik, Georg-August-Universit\"{a}t G\"{o}ttingen, Friedrich-Hund-Platz 1, 37077 G\"{o}ttingen, Germany.}
\altaffiltext{4}{Astronomy Department, New Mexico State University.}
\altaffiltext{5}{Visiting Astronomer, Kitt Peak National Observatory, National Optical Astronomy Observatory, which is operated by the Association of Universities for Research in Astronomy, Inc., under cooperative agreement with the National Science Foundation.}

\begin{abstract}

We use \textit{Hubble Space Telescope} Fine Guidance Sensor astrometry and high-cadence radial velocities for HD\,136118 from the HET with archival data from Lick to determine the complete set of orbital parameters for HD\,136118\,b.  We find an orbital inclination for the candidate exoplanet of $i_{b} = 163.1\degr\pm3.0\degr$. This establishes the actual mass of the object, $M_{b} = 42^{+11}_{-18} M_{J}$, in contrast to the minimum mass determined from the radial velocity data only, $M_{b}\sin{i} \sim12\,M_{J}$. Therefore, the low-mass companion to HD 136118 is now identified as a likely brown dwarf residing in the ``brown dwarf desert''.

\end{abstract}

\keywords{astrometry --- techniques: radial velocities --- planetary systems --- stars: low-mass, brown dwarfs --- stars: individual (HD136118) \newpage}

\section{INTRODUCTION}

\par Among the hundreds of exoplanets detected so far, only a few of them have their actual mass known. The widely used Doppler spectroscopy technique yields the radial component of the stellar perturbation velocity only. Consequently, the inclination of the orbital plane is unknown and only the minimum mass of a companion may be determined. To obtain a companion's true mass it is necessary to make use of additional techniques.

\par The first precise determination of an exoplanet mass was made for a transiting system \citep{henry2000}. However, transits are observed to occur only for systems that are oriented edge-on, and they have only a reasonable probability of occurrence for close-in planets (semimajor axes less than about 0.1\,AU). Another way to determine the orbital inclination of an unseen companion is by measuring the stellar reflex motion astrometrically. The first astrometrically determined mass of an exoplanet by \cite{benedict2002b} was possible thanks to the high precision of the Fine Guidance Sensor (FGS) instrument on the \textit{Hubble Space Telescope} (\textit{HST}). The FGS provides per observation precisions of better than $1$\,mas for small angle relative astrometry. This unique capability enables the detection of stellar perturbations due to planetary mass companions in wide orbits.

\par We were granted observing time with the \textit{HST} to measure the perturbation and determine the true mass of HD\,136118\,b, which is an exoplanet candidate found by radial velocity measurements \citep{fischer2002}.  To supplement previously published data and provide better constrain on the companion's spectroscopic orbital parameters we also obtained high-cadence radial velocity measurements with the Hobby-Eberly Telescope (HET). In this paper we present the results of our analysis and it is arranged as follows.  in Section \ref{sec:1} we review stellar properties for HD\,136118. In Section \ref{sec:2} we discuss the instrumental set up and data reduction for both spectroscopy and astrometry. In Section \ref{sec:3} we describe the orbital model used to analyze this data. In Section \ref{sec:4} we discuss the strategy employed to obtain the system parameters and present our results. In Section \ref{sec:5} we summarize and discuss the consequences of our results. 

\section{STELLAR PROPERTIES}
\label{sec:1}

\begin{figure}[!h]
\centering
\includegraphics[scale=0.6]{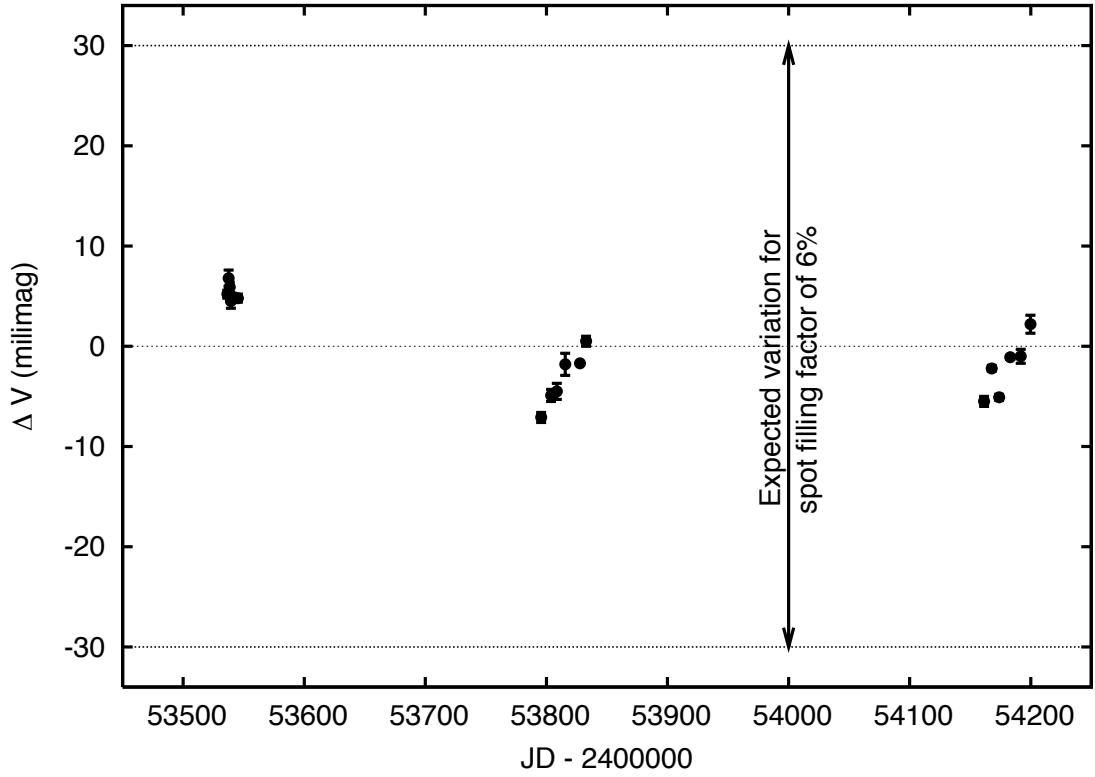}
\begin{footnotesize}
\caption{FGS-1r photometry of HD136118. Magnitude variation is relative to the mean magnitude, V=6.93. Dashed lines show
the amplitude of variation possible from a (single) spot filling factor of 6\%, the spot filling factor
required to produce the observed RV variation from HD136118 b \citep{saar1997}. \label{fig:hst_phot}} 
\end{footnotesize}
\end{figure}

HD\,136118 (=HIP\,74948) is a $V = 6.93$, F9\,V star with roughly solar photospheric abundances \citep{gonzalez2007}. Table \ref{tab:starpar} summarizes its observed properties given in the literature. Observations of Ca II H and K lines \citep{fischer2002} indicate modest chromospheric activity for this star, therefore not many spots should be expected. Using the \cite{saar1997} relationship for the spot radial velocity amplitude versus filling factor: $A_{S} = 6.5 f_{S}^{0.9} v\sin{i} $, where $f_{S}$ is spot filling factor in percent, $v\sin{i}$ is the projected velocity in km/s and $A_{S}$ is the spot radial velocity amplitude in m/s.  For the radial velocity amplitude of $211$\,m/s, and the measured velocity $v\sin{i} = 7\pm0.5$\,km/s  \citep{butler2006}, we obtain a spot filling factor of about 6\%, i.e. about 60 milimag variations. As shown in \cite{benedict1998} and \cite{nelan2007} the Fine Guidance Sensor itself is a millimag precision photometer. The variations we see in HD 136118 over 700 days are on order 4 parts per 1000, about 4 millimags, as shown on Fig \ref{fig:hst_phot}. This implies only small variations in spectral line shapes, which typically introduces noise on the order of $5-10$\,m/s in the velocities. 

\cite{fischer2002} have also provided evidence that the stellar rotation period is about $12.2$\,d. Given a stellar radius of $R = 1.58 \pm 0.11\,R_{\odot}$ \citep{allende1999} we calculate the maximum rotation speed at the stellar equator $v_{max} = 6.5\pm0.2$\,km/s. The measured projected velocity $v\sin{i} = 7\pm0.5$\,km/s is then consistent with the maximum speed. This suggests a very high inclination of the spin axis. Therefore, if the whole system follows the same angular momentum orientation as that of the star, then the companion's orbit would be close to an edge-on orientation with respect to our line of sight.

\section{OBSERVATIONS AND DATA REDUCTION}
\label{sec:2}

\subsection{HET Spectroscopy Data}
\label{sec:2-1}

Spectroscopic observations were carried out with the High Resolution Spectrograph \citep{tull1998} on the HET at the McDonald Observatory using the iodine absorption cell method \citep{butler1996}. Our observations include a total of 168 high resolution spectra which were obtained between UT dates 2005 December 4 and 2008 May 20.  Multiple observations were taken most nights, so the velocities obtained on the same night were combined, producing individual measurements of the stellar Radial Velocity (RV) at 61 different epochs.  

The spectrograph was used in the $R=60,000$ mode with a $316$\,line/mm echelle grating. The position of the cross dispersion grating was chosen so that the central wavelength of the order that fell in the break between the two CCD chips was $5936$\,\AA. A temperature controlled cell containing low pressure iodine (I$_{2}$) gas was placed in front of the spectrograph slit entrance during all the exposures. The absorption of light by the I$_{2}$ gas produces a set of well known spectral features imprinted at the same time as the stellar spectrum. However they are produced with no wavelength shift with respect to the observatory frame. This provides a much better reference for wavelength calibration and also permit us to characterize the instrumental profile with great accuracy. 
 
The exposure times were nominally $120$\,s, but were increased on a few nights due to bad seeing conditions. In addition to the program spectra we have also obtained HD\,136118 template spectra. For these we removed the I$_{2}$ cell, the resolution was set to $R=120,000$, and the exposure times were $230$\,s. 

A detailed description of our reduction and radial velocity analysis of HET HRS data is given in \cite{bean2007}. This provides us with measurements of the stellar radial velocity relative to an arbitrary zero point obtained from individual spectra. The offset of the entire data set is determined simultaneously with the orbital fit, then after compensating for this offset, we combined our RV data with previously published velocities from Lick Observatory \citep{fischer2002} to produce a total data set that spans $10.3$\,yr. Table \ref{tab:hetrvs} contains reduced HET data for the observed epochs.

\subsection{\textit{HST} Astrometry Data}
\label{sec:2-2}

Astrometric observations were obtained with the Fine Guidance Sensor 1r (FGS-1r), a two-axis, white-light interferometer aboard \textit{HST}. Our data were obtained with the FGS in fringe-tracking position mode. A detailed description of this instrument is found in \cite{nelan2007}.  All observations were secured under 2-gyro guiding, an operational mode dictated by gyro failures on \textit{HST}. This mode results in major constraints on \textit{HST} roll angle and observation dates. The dates of observation, the number of measurements for each date ($N_{obs}$) and the FGS-1r orientation angles are listed on Table \ref{tab:atmlog}. Our data sets span $1.8$\,yr, covering about 55\% of the companion's orbital period. 

\par In order to obtain high astrometric precision ($\sim 1$\,mas) with FGS we perform the following reduction procedures to remove the main sources of systematics. We apply an optical field angle distortion (OFAD) calibration  \citep{whipple1995, mcarthur1997}; apply time-dependent corrections to the OFAD \citep{mcarthur1997} with an additional, as yet unpublished improvement due to additional up-to-date M35 data that has provided better models to recalibrate the telescope; and we correct for drift during each observation set (intra-orbit drift). These procedures are described in detail in \cite{benedict1999,benedict2002c} and \cite{mcarthur2001}.

\par The FGS provides the measurement of each star position in a serial fashion. Each date listed in Table \ref{tab:atmlog} contains multiple measurements, alternating between the target (HD\,136118) and the reference stars (REF-14, REF-16 and REF-17), comprising a total of about 4 measurements per star. This provides $x'(t)$ and $y'(t)$ coordinates at a time $t$ of an epoch, which are in the \textit{HST} reference frame. 

\par A neutral density filter (F5ND) was applied when observing HD\,136118 due to its brightness. For the reference stars we used the F583W filter.

\subsection{Classification Spectra and Photometric Data}

\par We obtained classification spectra data for the astrometric reference stars. We include this in the models as bayesian a priori data to improve the accuracy of our determination of the orbital parameters for HD\,136118\,b. The spectra were obtained at the KPNO 4m Telescope and the photometric data at the NMSU 1m telescope in May of 2006. Table \ref{tab:ast1} summarizes spectral and photometric information for the reference stars.

\subsection{Astrometric Model}
\label{sec:2-3}

Given the positions ($x',y'$) measured by FGS-1r we build a model that accounts for positional changes occurring systematically in all reference stars from date-to-date. This is accomplished by solving an overlapping plate model which includes scaling-rotation (C$_1$, C$_2$, C$_3$, C$_4$) and offset ($D_1$,$D_2$) constants, which are constrained to an arbitrary frame adopted as the reference (the constrained plate). The astrometric model also accounts for the time dependent movements of each star, given by the absolute parallax  $\pi_{abs}$ and the proper motion components; $\mu_{\alpha}$ and $\mu_{\delta}$, where the indices specify the direction in equatorial coordinates. Therefore the model is given by the standard coordinates $\xi$ and $\eta$: 

\begin{eqnarray}
\xi & =  & C_1 x' + C_2 y' + D_1 - P_{\alpha}\pi_{abs} - \mu_{\alpha}\Delta t  \label{eq:astmodxi} \\
\eta &  = &  C_3 x' + C_4 y' + D_2 -  P_{\delta}\pi_{abs} - \mu_{\delta}\Delta t \label{eq:astmodeta}
\end{eqnarray}

where $P_{\alpha}$ and $P_{\delta}$ are parallax factors obtained from a JPL Earth orbit predictor (Standish 1990).  These are called the equations of condition, which comprises two equations for each star and for each epoch, giving a total of 144 equations to be solved simultaneously.  In order to find a global solution we used a program written in the GAUSSFIT language \citep{jefferys1988}. Table \ref{tab:ast2} contains the resulting astrometric catalog.

\cite{vanleeuwen2007} has recently presented a new reduction of \textit{Hipparcos} data. This resulted in an absolute parallax of HD\,136118 $\pi_{abs}=23.48\pm0.54$\,mas, which differs significantly from $\pi_{abs} = 19.13\pm0.85$\,mas, the prior \textit{Hipparcos} determination, and from our \textit{HST} result ($\pi_{abs} = 19.12\pm0.22$\,mas). When we constrained the parallax in our models to the new higher  \textit{Hipparcos} value,  we found that  the $\chi^{2}$ of our solution was increased by 20\%, which offers evidence that in this case, the older \textit{Hipparcos} determination was more accurate, if not more precise.

\section{ORBITAL MODEL}
\label{sec:3}

\subsection{Radial Velocity  Model}
\label{sec:3-1}

\par The velocity we are modeling is the radial component of the stellar orbital movement around the barycenter of the system, which is given by the projection of a Keplerian orbital velocity to observer's line of sight plus a constant velocity $\Gamma$. This constant in practice is not the actual velocity of the whole system but a parameter that absorbs all unaccounted for constants. Therefore we have the following equation:

\begin{equation}
v = \Gamma + K[\cos{(f+\omega)} + e \cos{\omega}]
\label{eq:rv}
\end{equation}      

where $\omega$ is the argument of periastron, $e$ is the eccentricity, $K$ is the velocity semi-amplitude and $f$ is the true anomaly. The latter gives the time dependence, which is obtained implicitly by solving the Kepler equation:

\begin{equation}
\frac{2\pi}{P}(t - T_{0}) = E - e\sin{E}
\end{equation}   

where $T_{0}$ is the epoch of periastron passage, $P$ is the orbital period and $E$ is the eccentric anomaly, which is related to $f$ by the following equation:
 
 \begin{equation}
\tan{\frac{f}{2}} = \sqrt{\frac{1+e}{1-e}} \tan{\frac{E}{2}}
\end{equation} 

The velocity semi-amplitude can also be written in terms of orbital elements:

\begin{equation}
K = \frac{2\pi}{P}\frac{a_{s}\sin{i}}{\sqrt{1 - e^2}}
\label{eq:velamp}
\end{equation} 

where $i$ is the orbital inclination and $a_{s}$ is the semi-major axis of the orbit of the star. Using the proportionality between the masses and semi-major axes, $a_{b}M_{b} = a_{s}M_{s}$, and Kepler's third law, we can rewrite Eq \ref{eq:velamp}:

\begin{equation}
\frac{M_{b}\sin{i}}{(M_{b} + M_{s})^{2/3}} = \left( \frac{P}{2\pi G} \right)^{1/3}K\sqrt{1-e^{2}}
\label{eq:massrv}
\end{equation} 

Note that we have introduced the indices $s$ and $b$ to distinguish between stellar and companion's parameters. Eq \ref{eq:massrv} provides a way to calculate the project minimum mass $M_{b}\sin{i}$ of the companion with the assumption that there is a measurement of the stellar mass by other means (e.g. stellar atmospheric models). It is to be noted that this is a lower limit to the mass with the uncertainties mostly dominated by the determination of the stellar mass.

\subsection{Apparent Orbit Model}
\label{sec:3-2}

Astrometric data provides positions of the parent star on the plane of the sky at different epochs. The high precision of FGS allows us to measure the apparent orbital movement of the star due to the presence of its companion. Therefore the astrometric observables to be modeled are the coordinates of the star apparent orbit. 
 
First we write the elliptical rectangular coordinates $x$,$y$, in the unit orbit, given by:

\begin{eqnarray}
x & = & (\cos{E} - e) \label{eq:xo}\\
y &  = & \sqrt{1-e^{2}} \sin{E} \label{eq:yo}
\end{eqnarray}

where $e$ is the eccentricity and $E$ is the eccentric anomaly. Remember that $E$ carries the dependence on time through the Kepler's equation. The projection of this true orbit onto the plane tangent to the sky gives the coordinates $\Delta x$,$\Delta y$. This projection can be expressed mathematically by:

\begin{eqnarray}			  	  
\Delta x & = & Bx + Gy \label{eq:x}\\
\Delta y & = & Ax + Fy \label{eq:y}
\end{eqnarray}

where $B,A,G,F$ are the Thiele-Innes constants, given by:

\begin{eqnarray}			  	  
B & = & \alpha (\cos{\omega} \sin{\Omega} + \sin{\omega}\cos{\Omega}  \cos{i}) \\
A & = & \alpha (\cos{\omega}\cos{\Omega}  -\sin{\omega} \sin{\Omega} \cos{i}) \\
G & = & \alpha (- \sin{\omega}\sin{\Omega} +\cos{\omega} \cos{\Omega} \cos{i}) \\
F & = & \alpha (-\sin{\omega}\cos{\Omega}  - \cos{\omega} \sin{\Omega} \cos{i}) 
\end{eqnarray}

 where $\alpha$ is the semi-major axis, $\Omega$ is the longitude of the ascending node, $i$ is the inclination of the orbit plane to the plane tangent to the sky and $\omega$ is the argument of periastron.  It should be noted that this can be the orbit coordinates for either the parent star or its companion around the barycenter, depending on which semi-major axis is taken; the $\omega$ in the respective orbits differ by $180\degr$. We measure the star orbit, so the coordinates of interest are $\Delta x_{s},\Delta y_{s}$,  obtained by taking $\alpha=\alpha_{s}$, which is expressed in mili-seconds of arc.  

\subsection{Combining RV and Astrometry}
\label{sec:3-2}

The orbital elements $(P, e, \omega, K)$ in the RV model are the same as the ones used in the astrometric model.  However, RV data has a time baseline much longer and a sampling much more abundant than that of the astrometric data. Therefore, in order to obtain the unknown parameters $\alpha_{s}$, $i$ and $\Omega$, we will use the RV parameters to enforce a ``constraint relationship'' between the astrometric and RV data sets.  A way to constrain the orbit without including the time dependence is by making use of Eq \ref{eq:velamp}. Rearranging the terms we have the following equation \citep{pourbaix2000}:

\begin{equation}
\alpha_{s}\sin{i}  =  \frac{KP\sqrt{1 - e^2}}{2\pi}
\label{eq:const}
\end{equation} 

where on the left side we have the projection of the semi-major axis of the perturbation orbit. The quantity $\alpha_{s}$ is an astrometric observable obtained in angular measure. It may be converted to linear measure (AU) by the relation $a [\tt{AU}] = \alpha/\pi_{abs}$. The right hand side carries all quantities obtained from RV.  

If parameters obtained from RV analysis are assumed as constants, there will be three independent measurements, Eqs \ref{eq:x}, \ref{eq:y} and \ref{eq:const}, to determine only three parameters, $\alpha_{s}$, $i$ and $\Omega$. It means that even if one has a poor astrometric coverage of the orbit, it would still converge to a unique solution, although uncertainties in the resulting orbital parameters could be reduced with additional coverage.

The final solution is obtained by performing a simultaneous fit of Eqs \ref{eq:rv} [$v(P,T,e,\omega,K,\Gamma;t)$], \ref{eq:x} [$\Delta x_{s}(P,T,\alpha_s,i,\omega,\Omega;t)$] and \ref{eq:y} [$\Delta y_{s}(P,T,\alpha_s,i,\omega,\Omega;t)$], for RV and astrometry data. Some parameters are constrained via Eq \ref{eq:const} and all parameters are in some way constrained by their physical meaning.

\section{RESULTS}
\label{sec:4}

All solutions presented below were obtained using GAUSSFIT \citep{jefferys1988} programs that minimize $\chi^{2}$ for the equations shown in the models. As in Section \ref{sec:3-1} above index $b$ stands for the companion and $s$ for the parent star. 
    
\subsection{Radial Velocity Solution}
\label{sec:4-1}

\begin{figure}[!h]
\centering
\includegraphics[scale=0.6]{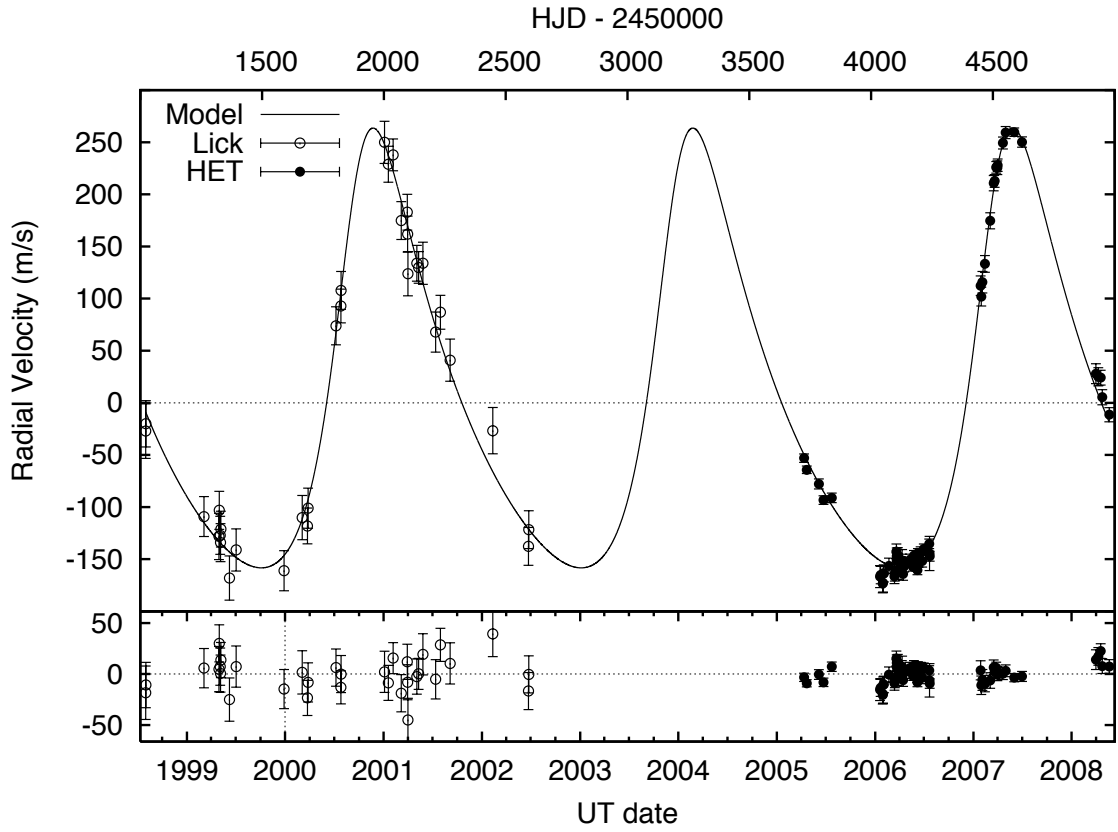}
\begin{footnotesize}
\caption{HD\,136118 RV HET (filled circles) and Lick (open circles) data and the best fit model (solid line). \label{fig:rvplot-time}} 
\end{footnotesize}
\end{figure}

The best fit model solution yields to orbital parameters shown on Table \ref{tab:rvparms}.  Fig \ref{fig:rvplot-time}  shows HET RV data plotted together with previous published Lick data \citep{fischer2002}. The fitted model is also plotted. The residuals are shown in the bottom panel. Table \ref{tab:sq} shows the statistical quantities obtained from the residuals, where we used the following terminology.  Central values: mean and median. Standard Deviation: $\sigma$. Sum of squared normalized residuals: $\chi^{2}$. Degrees of Freedom: $\tt{DOF}$. Reduced chi-square: $\chi^{2}_{\nu} = \chi^{2}/\tt{DOF} $.  We estimate these quantities for each data set separately and also for all data sets combined together.  These allow us to evaluate the legitimacy of the fit.

\begin{figure}[!h]
\centering
\includegraphics[scale=0.5]{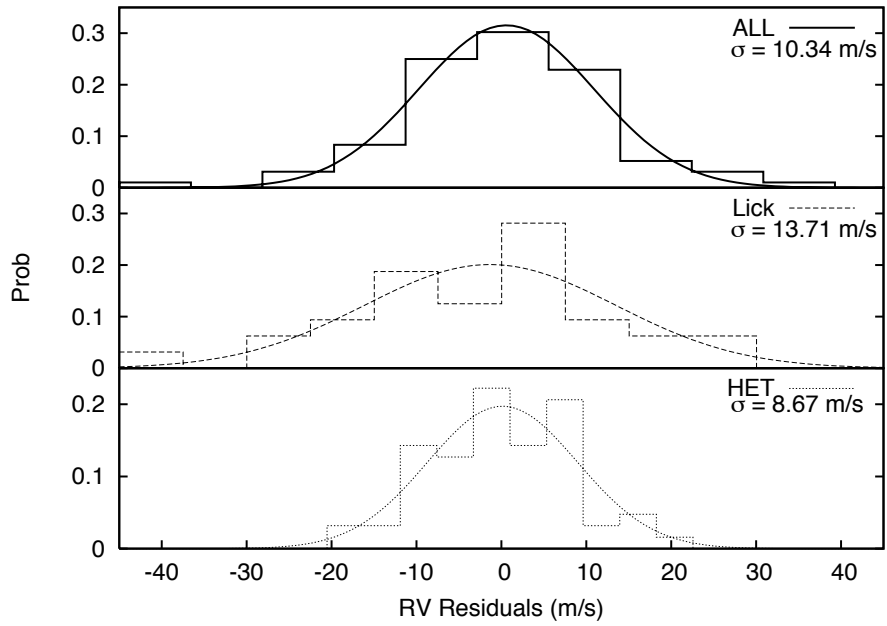}
\begin{footnotesize}
\caption{Histogram of residuals from 1-companion model for 3 datasets: all combined (ALL) (top panel), Lick (middle panel) and HET (bottom panel). \label{fig:hist-res}} 
\end{footnotesize}
\end{figure}

From Table \ref{tab:sq} we note that $\chi^{2}_{\nu}$ for Lick data is close to unity, indicating satisfactory agreement between the dispersion of residuals and the individual errors. However this agreement between dispersion and errors is not as definitive for the HET data.  We firstly check whether the residuals follow a Gaussian distribution and inspect the errors involved. Fig \ref{fig:hist-res} shows the histogram of distribution of residuals for the two individual datasets separately, Lick and HET,  and also for both datasets combined (hereafter ``ALL'').  We can see the different dispersion for each dataset. We note that each individual dataset, either HET and Lick, are not exactly following a normal distribution.  We call attention to the fact that the dispersion on HET residuals is about three times larger than the error ($\sim3$\,m/s) estimated from previous work [e.g. \cite{bean2007}]. This discrepancy may be identified with an unaccounted for systematic effect.  Below we investigate the detection limits for any further periodic signal that could still be present in our data.

\subsection{Limits on Additional Periodic Signals in the RV Data}
\label{sec:4-2}

\begin{figure}[!h]
\centering
\includegraphics[scale=0.5]{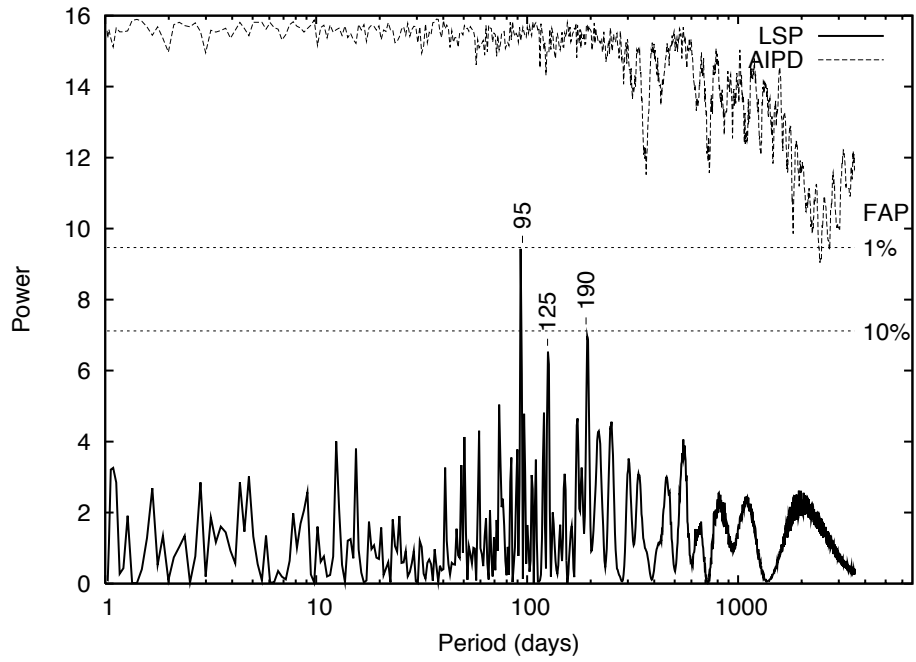}
\begin{footnotesize}
\caption{ Lomb-Scargle periodogram for RV residuals from 1-companion model and using all datasets combined (solid line). The thresholds for false alarm probability of 1\% and 10\% are plotted in dotted lines.  Above it is shown the AIPD (see text) for the same dataset (dashed line). The AIPD here is multiplied by a factor 20 and shifted for the sake of better visualization. \label{fig:lsp_2ds}} 
\end{footnotesize}
\end{figure}

\begin{figure}[!h]
\centering
\includegraphics[scale=0.5]{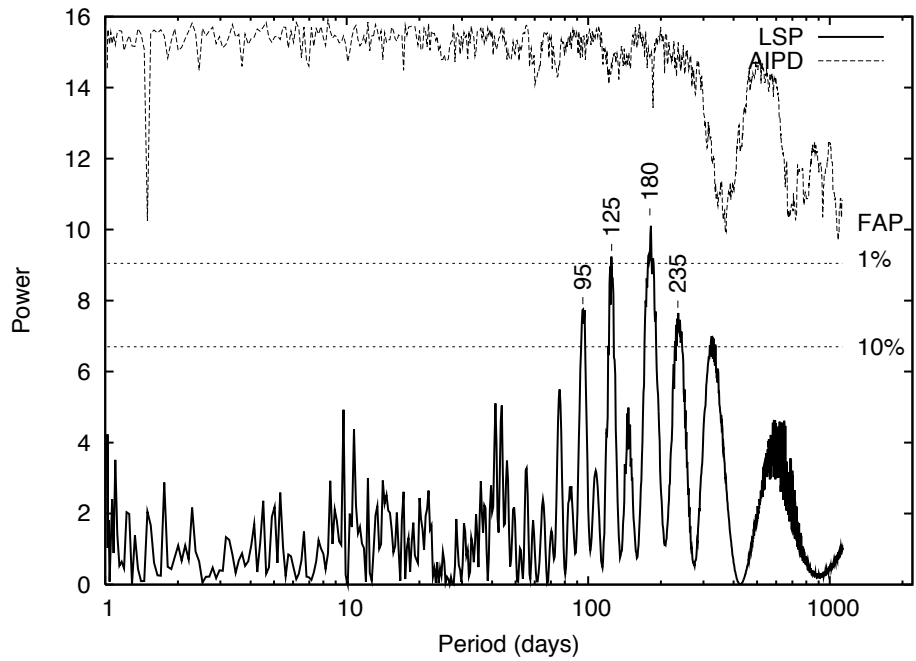}
\begin{footnotesize}
\caption{ Lomb-Scargle periodogram for HET RV residuals from 1-companion model (solid line). The thresholds for false alarm probability of 1\% and 10\% are plotted in dotted lines.  Above it is shown the AIPD (see text) for the same dataset (dashed line). The AIPD here is multiplied by a factor 20 and shifted for the sake of better visualization. \label{fig:lsp_het}} 
\end{footnotesize}
\end{figure}

The customary method for searching periodic signals in unevenly spaced data is by means of the Lomb-Scargle Periodogram (LSP) \citep{scargle1982}. Figs \ref{fig:lsp_2ds} and \ref{fig:lsp_het} show the LSP of residual RV data for two different datasets respectively: ALL and HET. The sets are analyzed separately because they have different errors (see Table \ref{tab:rvparms}). The power in the LSP is weighted by the overall variance, therefore if one mixes two sets with different variances it would result in an overestimated power for higher levels of noise. The downside of analyzing datasets separately is that sampling becomes different as you have different time coverage and it may affect the detectability for some frequencies.

The LSP for Lick data does not seem to show any expressive power \citep{fischer2002}. The LSP for HET dataset (Fig \ref{fig:lsp_het}) shows some peaks at the limit where False Alarm Probability (FAP) is as low as about 1\%. The combined dataset also shows some power below the level of 1\% FAP. This indicates that either there is still an unaccounted periodic signal or the sampling for those frequencies is poor. The latter may be analyzed by the method we describe below.

We introduce a quantity to evaluate how much we can trust some high power found at a given frequency based on sampling for that frequency. We call this quantity the Amount of Information in the Phase Diagram (AIPD). It is defined by the following expression: 

\begin{equation}
I = - \sum_{i=1}^{N_b}{p_i\ln{p_i}}/\ln{N_b}
\label{eq:aipd}
\end{equation}

where $N_b$ is the number of bins in the phase diagram, $p_i$ is the probability of finding a data point within a given bin $i$, and may be calculated by $p_i = n_i/N$, where $n_i$ is the number of points inside the bin $i$ and $N$ is the total number of data points.  Note that the dependence on the period arises from the construction of the phase diagram.  This quantity is normalized and therefore it varies from 0 to 1. When $I=0$ it means that all data points are found within a single bin in the phase diagram and hence sampling for that frequency is very poor. From the other hand when $I=1$ it means that probability $p_i$ is the same for every bin and data is equally distributed along all bins. This gives you an ideal coverage of the phase diagram. We call attention to the fact that AIPD does not measure the statistical significance of the number of data points but only the significance of how well distributed are these points in the phase diagram.  Therefore an issue of concern is the choice of an adequate $N_b$.  We suggest the choice is made in the same fashion as when you build a traditional histogram for inspecting probability distributions. Our choice of $N_b$ is that of $N/N_b > 30$\,data-points-per-bin. If $N$ is small such that makes $N_b < 10$ then we assume $N_b=10$.

Figs \ref{fig:lsp_2ds} and \ref{fig:lsp_het} also show a plot of AIPD (in the plot it is multiplied by 20 and shifted for the sake of visualization). We note that some of the high power periods in the LSP also presents a decreasing on the AIPD, which means a deficit of sampling for those periods. This is evident for the 1\,yr period where there is always lack of data for some part of the phase diagram. If you fold the 1yr phase diagram twice there will still be some lack of data coming from the 1\,yr sampling problem. This can be seen from the smaller decreases at half year period. Although it is smaller it may still affect the LSP. If one disregards the powers at periods which are close integer fractions of a year there will be no significant power left on the LSPs.

However, the existence of an additional signal cannot be ruled out by looking only at the LSP for the following reason.  If the signal comes from an orbit which follows the RV model (Eq \ref{eq:rv}), then it may be considered solutions for eccentric orbits instead of strict sine and cosines as in the LSP.  An alternative periodogram using the orbit solution is explored in \cite{gregory2007}. We propose a strategy to find a possible hidden signal in our particular case, although it could be expanded and applied for any other system. 

\begin{figure}[!h]
\centering
\includegraphics[scale=0.4]{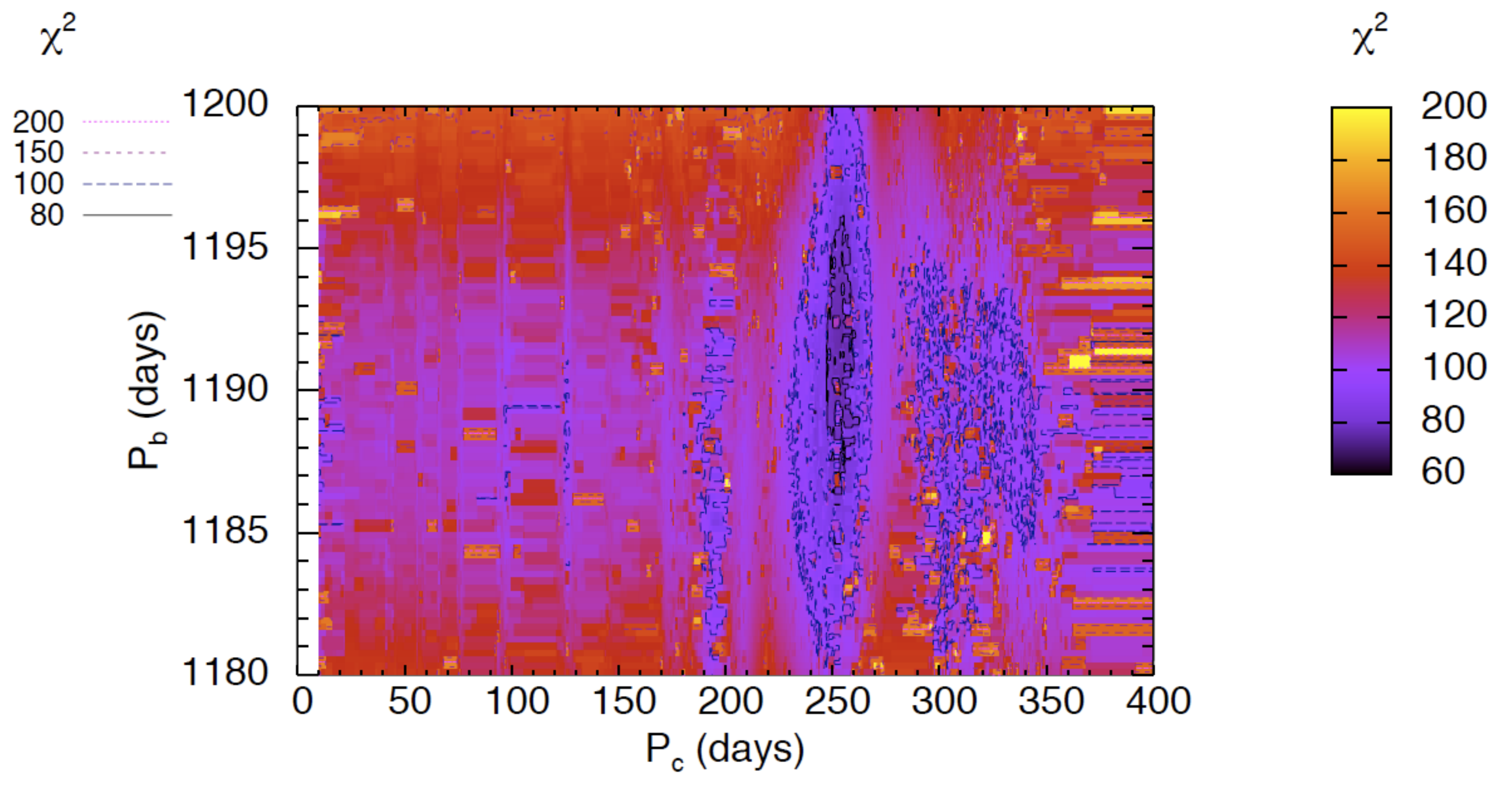}
\begin{footnotesize}
\caption{$\chi^2$ map from a 2-companion fit model for HET and Lick data. The grid resolution is about $0.3$\,day ($1100\times90$ points).  Contour lines show four different levels of $\chi^2$. The best fit solution has the lowest value at $\chi^2 = 76.6$.  \label{fig:chi2}}
\end{footnotesize}
\end{figure}

Our strategy consists in making multiple attempts to model the RV, using a 2-companion model (linear superposition of two Keplerian orbits), and keeping the trial periods as constants in the fitting process. This process forces the minimization algorithm to search for the best solutions for each chosen period. This approach could result in a lower $\chi^2$ when including a hidden periodic component. Fig \ref{fig:chi2}  shows a $\chi^2$ map over a range of periods for a 2-companion model fitting RV HET and Lick data simultaneously. The grid point resolution is $1100\times90$, which means a step on the trial periods of about $\sim 0.3$\,day for both components. From Fig \ref{fig:chi2} we can see a region around 255\,days where we found an island of lower $\chi^2$ (darker regions), which indicates the presence of an additional signal. We note that the relatively high power peak at $\sim95$\,days in the LSP (see Fig \ref{fig:lsp_2ds}) is now ruled out, because any attempt of fitting a secondary orbit with this period results in larger $\chi^2$. This approach does not prove the existence of another companion in the system.  Rather it shows an effective way of finding solutions that are considerably improved by adding a weak periodic signal that could not be detected in the LSP. Moreover, our model uses Keplerian orbits, which look not only for a periodic signal but for a signal with the shape of an orbit.

\begin{figure}[!h]
\centering
\includegraphics[scale=0.5]{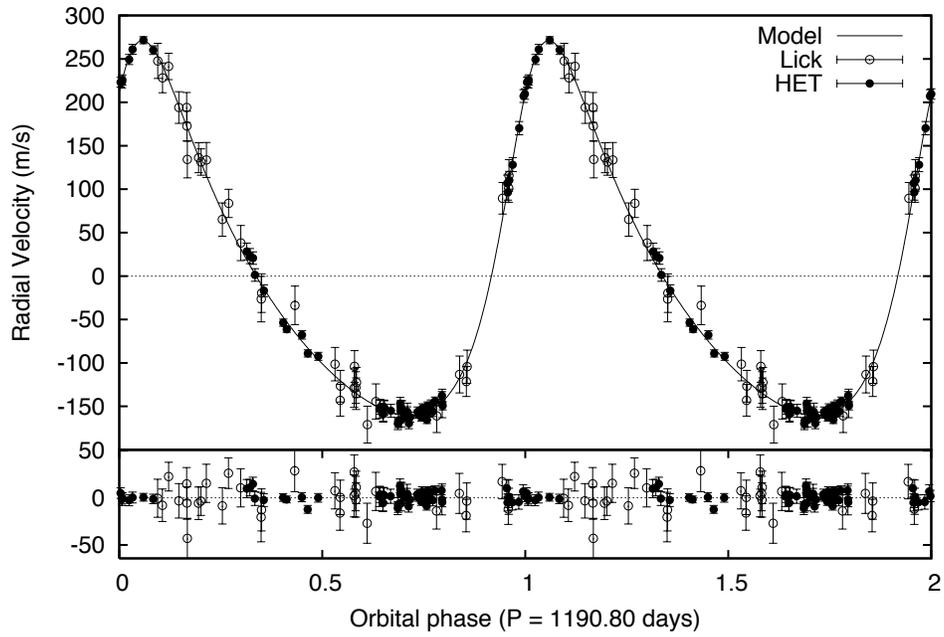}
\begin{footnotesize}
\caption{Phase diagram folded with period 1190.8\,days. RV HET (filled circles) and Lick (open circles) data subtracted the ``nuisance orbit'' model.  Solid line shows the best fit HD\,136118\,b orbit model. Residuals from the 2-companion model is plotted in the bottom panel. \label{fig:phase_b}} 
\end{footnotesize}
\end{figure}

\begin{figure}[!h]
\centering
\includegraphics[scale=0.5]{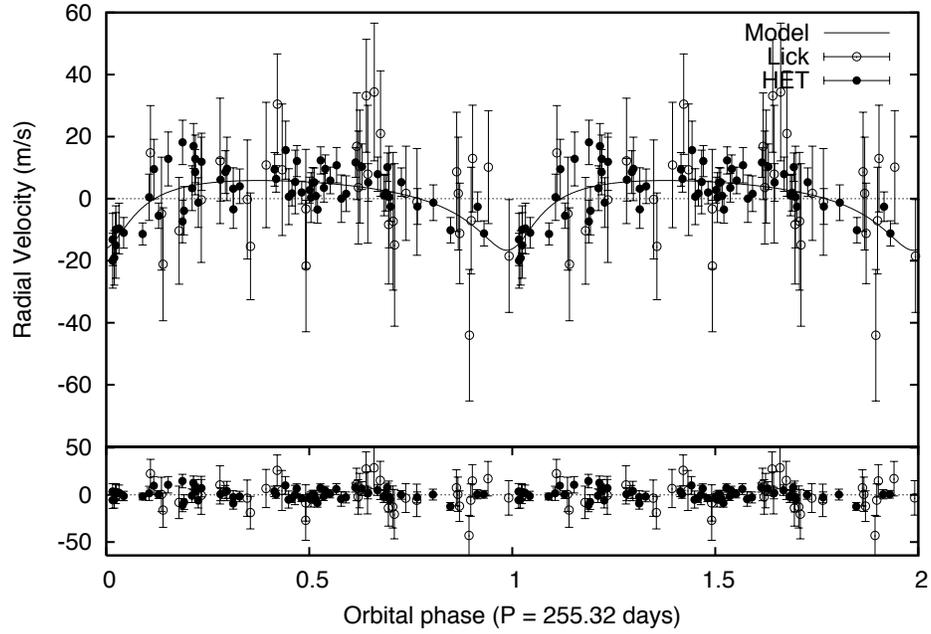}
\begin{footnotesize}
\caption{RV HET (filled circles) and Lick (open circles) residuals from HD\,131168\,b orbit model and the best fit ``nuisance orbit'' model plotted in the phase diagram folded with period 255.32\,days.  Residuals from the 2-companion model is plotted in the bottom panel. \label{fig:phase_c}} 
\end{footnotesize}
\end{figure}

Then, we performed a refined fit using the Levemberg-Marquardt method and then a robust fit method for a 2-companion model with an additional signal with period of about 255 days. This solution presents a notable improvement on the minimization of $\chi^2$ if compared to the 1-companion model (see Tables \ref{tab:sq} and \ref{tab:sq-2p}). The parameters for a 2-companion model are shown on Table \ref{tab:rvparms-2p}.  From Table \ref{tab:sq-2p} we note an improvement on all statistical quantities. The $\sigma$ dispersion for HET data is now in agreement with that we expected. Figs \ref{fig:phase_b}  and \ref{fig:phase_c} show the phase diagram of RV data and the respective component orbit model. 

Such solution suggests the presence of an additional lower mass companion to the system. However this would be a very eccentric orbit and considerably large $M\sin{i}$ planet, which makes us believe that such body could hardly coexist with a brown dwarf orbiting the same system at the distances they seem to be. In fact we have performed a stability analysis using the MERCURY package \citep{chambers1999}.  We have input the two companions around HD\,136118, and considered the minimum mass, which minimizes interactions. We also explored the full range of inclinations for the second component.  The system always becomes unstable for very short time scales.  Besides the stability constraint, this detection is at the limit of our instrument, and also if we look at Fig \ref{fig:phase_c} we note that only a few data points are contributing to form the orbit we have obtained. For these reasons we prefer to be cautious and take this as a ``nuisance orbit'' to fix an unknown source of systematic error present in the HET data, although the possibility of a second companion is not out of the question. Further re-reductions and observations will be done soon to investigate the origin of this signal.  The parameters adopted in the following sections are those from a 2-companion model. The modelling of the second component in the RV has a very marginal effect on the parameters of the astrometric detection, so it is not ``polluting'' our result.

\subsection{Simultaneous RV and Astrometry Solution}
\label{sec:4-3}
 
\begin{figure}[h!]
 \centering
\begin{minipage}[c]{1\textwidth}
   \centering \includegraphics[scale=0.9]{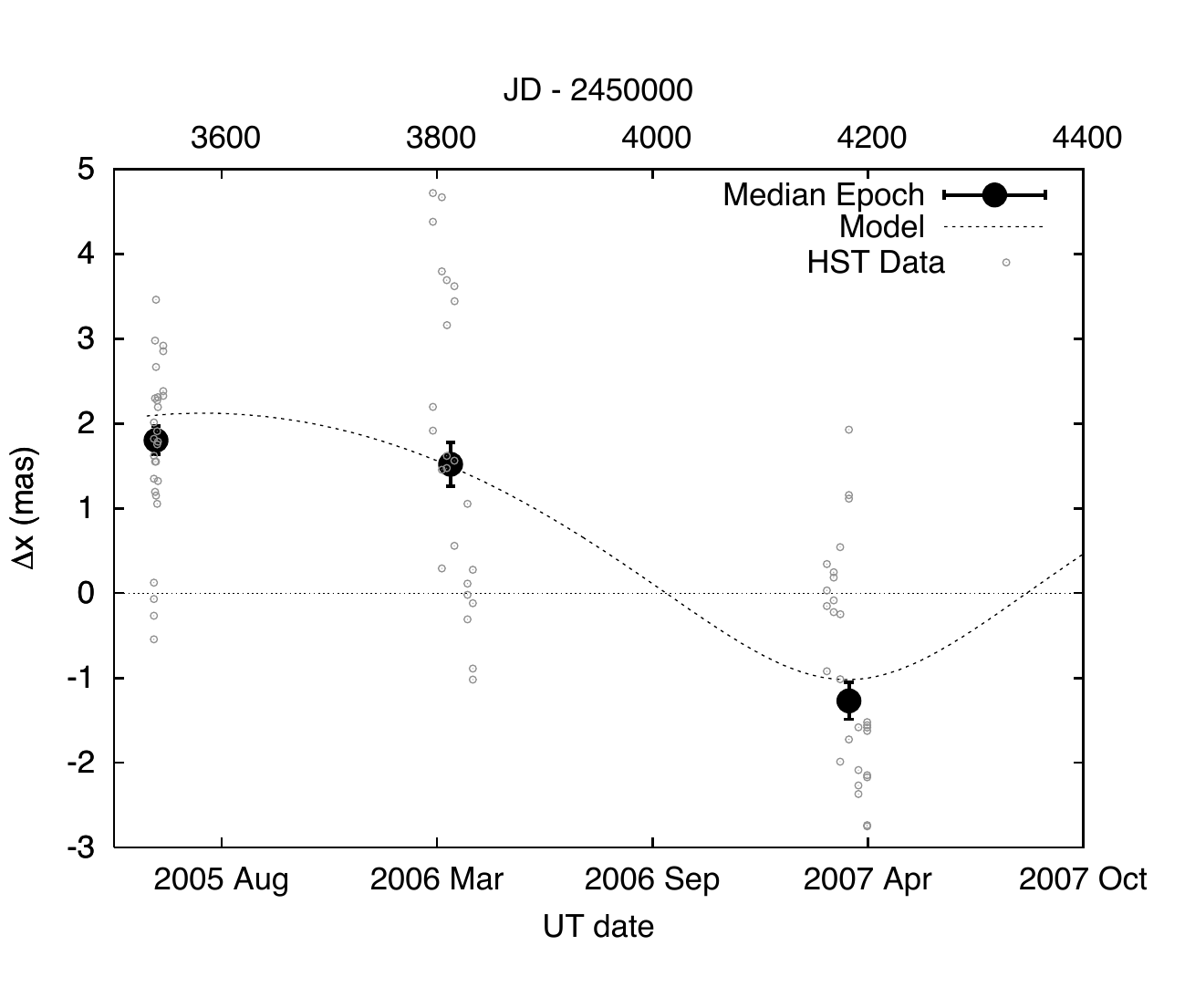}
 \end{minipage}%

 \begin{minipage}[c]{1\textwidth}
    \centering \includegraphics[scale=0.9]{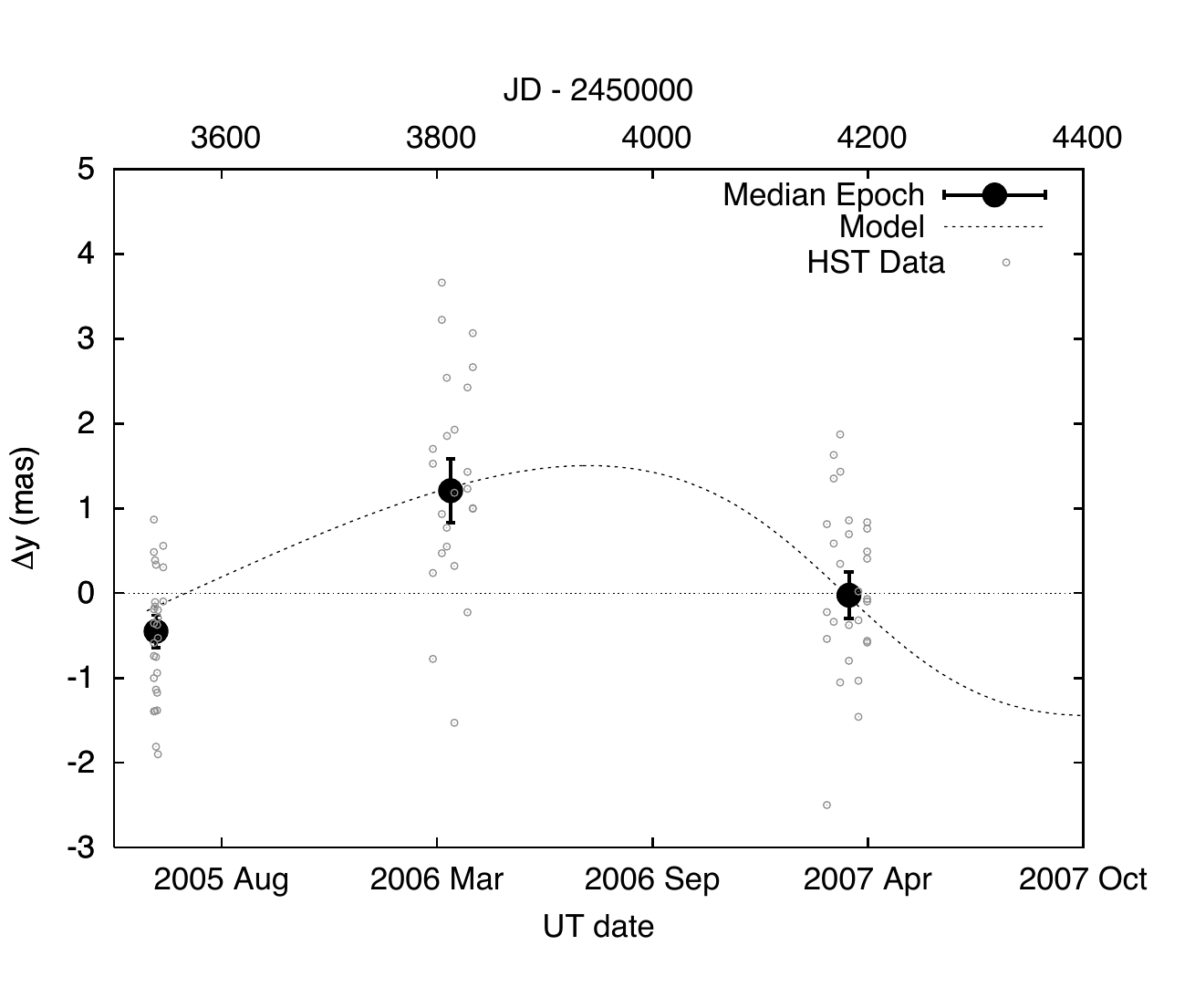}
 \end{minipage}
 \caption{Open circles are the astrometric residuals of HD\,136118 in $\Delta x$ (upper panel) and $\Delta y$ (bottom panel) versus time. Filled circles with error bars represent 3 epochs. These were obtained through the median and standard deviation of each grouped data.  The perturbation orbit fit model is plotted in dashed lines.\label{fig:astrxyt}}
\end{figure}

\begin{figure}[!h]
\centering
\includegraphics[scale=1.2]{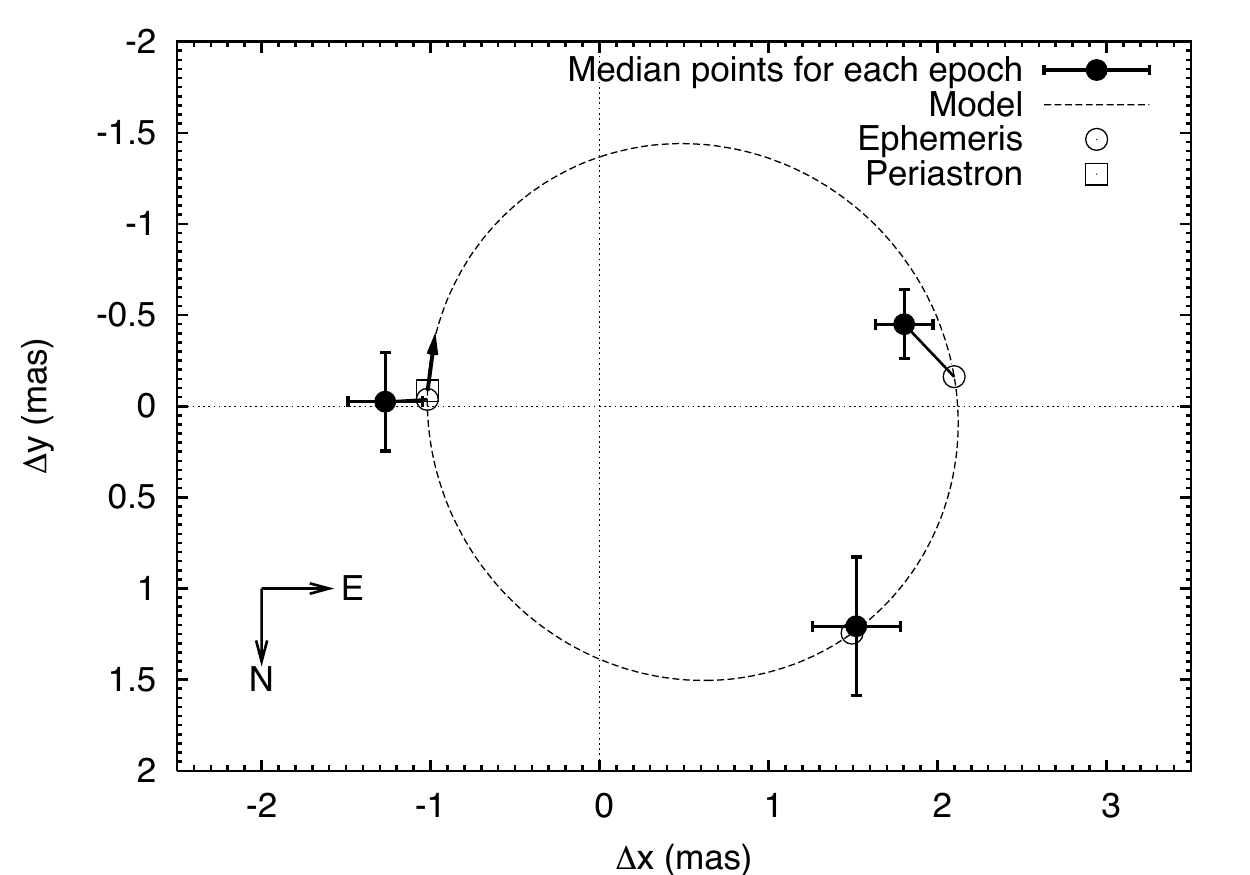}
\begin{footnotesize}
\caption{Filled circles with error bars are the median and standard deviation of three groups  of astrometric residuals of HD\,136118, representing three different epochs. Dashed line is the fit model of the apparent perturbation orbit of HD\,136118.  Open circles are the positions calculated from the fit model, each of which is connected by a solid line to its respective observed epoch. The open square shows the predicted position of the periastron passage. \label{fig:astrxy}} 
\end{footnotesize}
\end{figure}

\begin{figure}[!h]
\centering
\includegraphics[scale=0.5]{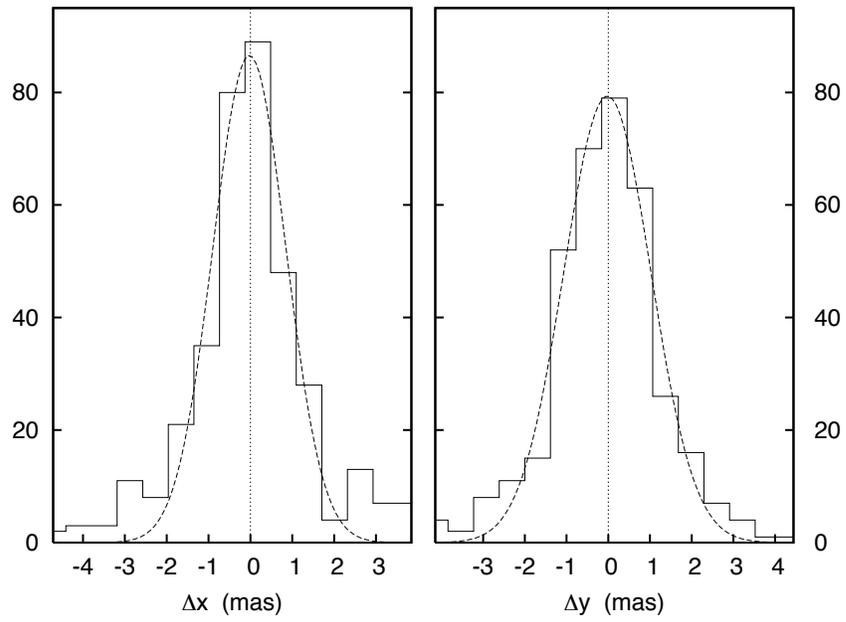}
\begin{footnotesize}
\caption{ Histogram of $\Delta x$ (left panel) and $\Delta y$ (right panel) astrometric residuals for all reference stars. Dashed lines show the fit model for a normal distribution.  \label{fig:hst_res}} 
\end{footnotesize}
\end{figure}

The system parameters obtained in Section \ref{sec:4} from RV analysis are intially  adopted as constants as we search for orbital solutions in the astrometric data.  Once close to the solution we free all parameters and obtain the best fit model for astrometry and radial velocity simultaneously. We obtain a semi-major axis of the perturbation orbit $\alpha_{s} = 1.45\pm0.25$\,mas, an inclination $i = 163.1\degr\pm3.0\degr$ and a longitude of the ascending node $\Omega =  285\degr\pm10\degr$. Fig \ref{fig:astrxyt} shows the reduced star positions $\Delta x_{s}$ and $\Delta y_{s}$, versus time for HD\,136118. Although our solutions were obtained considering each data point individually, in Fig \ref{fig:astrxyt}, in order to provide the reader a better visualization to show how the fit works,  we also plot normal points which are the median and respective standard deviation of the mean of each clump of data, representing 3 different epochs. These collapsed points are also shown on Fig \ref{fig:astrxy}, where we plotted $\Delta y_{s}$ versus $\Delta x_{s}$ and the apparent orbit fit model. Fig \ref{fig:hst_res} shows the distribution of astrometric residuals of $\Delta x$ (left panel) and $\Delta y$ (right panel) for all reference stars and all data sets. A Gaussian fit model is superposed for comparison. The fit distribution for both $\Delta x$ and $\Delta y$ present a maximum consistent with zero, and a FWHM of $0.87$\,mas and $1.02$\,mas respectively.

\par By determining the inclination we are able to remove the previous degeneracy on the mass of the companion. We calculate the actual mass by iterating Eq \ref{eq:massrv}, which yields $M_{b} = 42^{+11}_{-18} \,M_{J}$, firmly establishing HD\,136118\,b as a bloom in the brown dwarf desert. We also obtain the physical semi-major axis of the companion orbit, $a_{b} = 2.36\pm0.05$\,AU. A summary containing all HD\,136118\,b parameters derived from the simultaneous RV and astrometry solution is shown on Table \ref{tab:allpars}.

\section{SUMMARY AND DISCUSSION}
\label{sec:5}

HD\,136118 is a solar type star with a brown-dwarf companion. Table \ref{tab:allpars} shows a summary of all observed orbital elements of the system. 

We found that HD\,136118\,b has an orbital inclination of $i = 163.1\degr\pm3.0\degr$, nearly perpendicular to the inferred inclination of the stellar spin axis. This misalignment is an intriguing result since conservation of angular momentum would favor alignment between stellar spin and companion orbital axes, assuming both were formed in the same primordial cloud. 

HD\,136118\,b is likely a brown dwarf companion orbiting at $2.36\,$AU that falls in the driest region of the so called `brown dwarf desert' \citep{grether2006}.  They showed that the frequency of companions in the stellar mass range follows a slope with gradient $-9.1\pm2.9$, while in the planetary mass region, the gradient is $24.1\pm4.7$. These two separate linear fits intersect below the abscissa at $M=43^{+14}_{-23}\,M_{J}$. Surprisingly, HD\,136118\,b mass is  $M_{b}=43^{+11}_{-18}\,M_{J}$. \cite{reffert2006} measured  astrometric masses for two exoplanet candidates HD\,38529\,c ($M = 37^{+36}_{-19}\,M_{Jup}$) and HD\,168443\,c ($M = 34\pm12\,M_{Jup}$). Both are likely brown dwarf companions around solar type stars like HD\,136118\,b. These objects are important cases for studying the mass function at the brown dwarf mass range.

\begin{figure}[!h]
\centering
\includegraphics[scale=0.6]{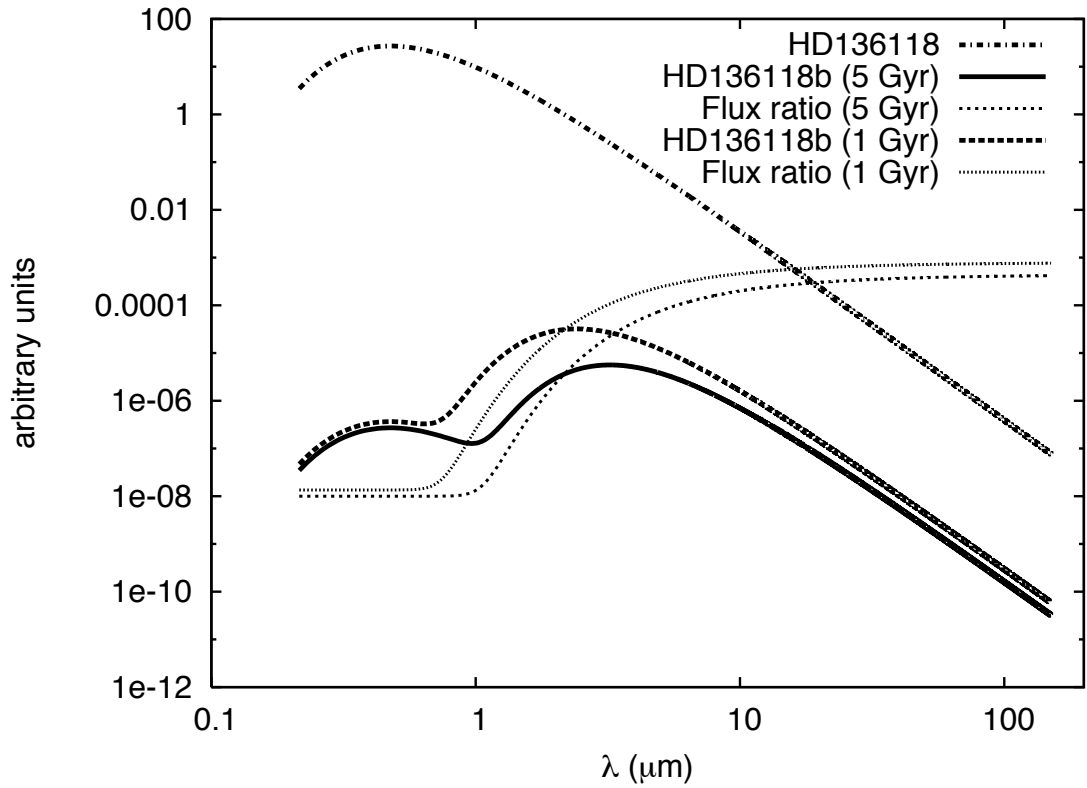} 
\begin{footnotesize}
\caption{Predicted emission spectrum of HD\,136118 (dash-dotted line),  emission/reflection spectrum of HD\,136118\,b and the flux ratio between the brown dwarf and the parent star for a 5\,Gyr and 1\,Gyr system age, as indicated in the legend. \label{fig:flux-ratio}} 
\end{footnotesize}
\end{figure}

According to the evolutionary dusty model for brown dwarfs of \cite{baraffe2001} and \cite{chabrier2000}, assuming $M_{b} = 0.041^{+0.010}_{-0.017}\,M_{\odot}$  and the age of the brown dwarf as $5$\,Gyr, HD\,136118\,b has a temperature of about $T_{b} = 900$\,K and a radius of $R_{b} = 0.086\,R_{\odot}$. If one considers the uncertainty in the age of the system (see Table \ref{tab:starpar}), this brown dwarf may be much younger, therefore considering the age as $1$\,Gyr, HD\,13118\,b has a temperature of about $T_{b} = 1200$\,K and  a radius of $R_{b} = 0.1\,R_{\odot}$. These characteristics classifies HD\,136118\,b as a T dwarf. Using these values we calculate the emission and reflection spectra, and the flux ratio between the brown dwarf and the parent star as show in Fig \ref{fig:flux-ratio}. The flux ratio increases toward the far infrared (L, M and N bands), where it can get as high as $10^{-4}$.

The astrometric determination of the mass of a low mass companion can decisively characterize it as a planet.  A good illustration of this fact can be seen from the results of our group for three objects that were previously listed as exoplanet candidates: Gliese\,876\,b, HD\,136118\,b and HD\,33636\,b.  Surprisingly each object has been found to belong to a different class: a giant planet, a brown dwarf and a M dwarf star, respectively, \cite{benedict2002b}, this paper, and \cite{bean2007}. These results demonstrate the importance of the application of complementary techniques in observing extrasolar planetary systems.

\acknowledgments

This work was supported by the {\it Coordena\c{c}\~ao de Aperfei\c{c}oamento de Pessoal de N\'ivel Superior} (CAPES), Brazil. Support for this work was also provided by NASA through grants GO-10704-10989, and-11210 from the Space Telescope Science Institute, which is operated by AURA, Inc., under NASA contract NAS5-26555. This research is based partially on observations carried out with the Hobby-Ebberly Telescope at McDonald Observatory. This research has made use of the SIMBAD database, operated at  CDS, Strasbourg, France;  and NASAs Astrophysics Data System Abstract Service.

{\it Facilities:} \facility{HET(HRS)}, \facility{\textit{HST}}.



\clearpage

\begin{deluxetable}{lccc}
\tablewidth{0in}
\tablecaption{Properties of HD\,136118.\label{tab:starpar}}
\tabletypesize{\scriptsize}
\tablehead{\colhead{ID}&\colhead{HD\,136118}&\colhead{unit}&\colhead{ref}}
\startdata

RA(2000) & $15:18:55.4719\,(8.18) $  &h:m:s& \tablenotemark{a} \nl
Dec(2000) & $ -01:35:32.590\,(5.37) $  &d:m:s&  \tablenotemark{a} \nl
$\mu_{\alpha}$& $ -124.1\,(0.9) $  &mas\,yr$^{-1}$ &  \tablenotemark{h} \nl
$\mu_{\delta}$& $23.5\,(0.7) $  &mas\,yr$^{-1}$ &  \tablenotemark{h} \nl
$\pi_{abs}$ & $19.1\,(0.8)$ & mas  &   \tablenotemark{h}  \nl
$\Gamma$& $-3.6\,(0.1) $  & km\,s$^{-1}$ &  \tablenotemark{a} \nl
Spc type & F9V & - &  \tablenotemark{h}  \nl
Age  & 4.8 $(^{+0.7}_{-1.9})$& Gyr &   \tablenotemark{e}  \nl
 $[$Fe/H$]$  & -0.010\,(0.053) & dex &  \tablenotemark{g}   \nl
 $[$C/H$]$  & 0.049\,(0.081) & dex &  \tablenotemark{g}   \nl
 $[$O/H$]$  & 0.112\,(0.045) & dex &  \tablenotemark{g}   \nl
 $[$Si/H$]$  & -0.042\,(0.058) & dex &  \tablenotemark{g}   \nl
 $[$Ca/H$]$  & -0.057\,(0.062) & dex &  \tablenotemark{g}   \nl
$d$  & 52.3\,(0.6) & pc &   \tablenotemark{h}  \nl
$v\sin{i}$  & 7.33\,(0.5) & m\,s$^{-1}$ &  \tablenotemark{f}   \nl
$P_{rot}$  & 12.2 & day &  \tablenotemark{c}   \nl
$T_{eff}$  & 6097\,(44)  & K &   \tablenotemark{f} \nl
$\log{g}$  & 4.16\,(0.09) & cm\,s$^{-2}$ &  \tablenotemark{b}   \nl
$M_{*}$ & 1.24\,(0.07) & M$_{\odot}$ &  \tablenotemark{c}   \nl
$R_{*}$   & 1.58\,(0.11)& R$_{\odot}$ &   \tablenotemark{b}  \nl
BC & 0.01\,(0.03) & mag &  \tablenotemark{b}  \nl
$M_{V}$ & 3.34 & mag &  \tablenotemark{d}  \nl
$B$  & 7.432  & mag &  \tablenotemark{d}  \nl
$V$  & 6.945 & mag &  \tablenotemark{d}  \nl
$R$  & 6.630 & mag &  \tablenotemark{d}  \nl
$J$  & 5.934 & mag &  \tablenotemark{d}  \nl
$H$  & 5.693 & mag &  \tablenotemark{d}  \nl
\enddata

\tablenotetext{a}{\cite{perryman1997}}\tablenotetext{b}{\cite{allende1999}}\tablenotetext{c}{\cite{fischer2002}} \tablenotetext{d}{\cite{zacharias2004}}\tablenotetext{e}{Age value and limits derived from isochrone method \citep{saffe2005}}\tablenotetext{f}{\cite{butler2006}}\tablenotetext{g}{\cite{gonzalez2007}} \tablenotetext{h}{This paper} 
\end{deluxetable}

\begin{deluxetable}{ccc}
\tablewidth{0in}
\tablecaption{HET relative radial velocities for HD\,136118.\label{tab:hetrvs}}
\tabletypesize{\scriptsize}
\tablehead{\colhead{HJD - 2450000}&\colhead{RV (m/s) }&\colhead{$\pm$ error}}
\startdata
3472.831	&	432.6	&	4.1	\nl
3482.881	&	421.4	&	3.5	\nl
3527.763	&	407.8	&	4.8	\nl
3544.727	&	392.4	&	4.1	\nl
3575.630	&	394.6	&	4.6	\nl
3755.051	&	319.0	&	10.6	\nl
3757.041	&	320.5	&	8.4	\nl
3765.026	&	312.2	&	8.8	\nl
3766.026	&	313.0	&	8.8	\nl
3767.020	&	321.9	&	7.7	\nl
3769.011	&	321.9	&	8.2	\nl
3787.982	&	329.2	&	7.5	\nl
3808.904	&	319.1	&	6.9	\nl
3809.909	&	322.5	&	7.8	\nl
3815.886	&	342.9	&	7.3	\nl
3816.898	&	334.4	&	7.3	\nl
3816.965	&	338.7	&	7.7	\nl
3818.873	&	324.5	&	8.0	\nl
3820.897	&	337.5	&	9.2	\nl
3832.840	&	331.2	&	6.9	\nl
3835.853	&	333.6	&	6.9	\nl
3836.858	&	334.6	&	10.2	\nl
3840.895	&	321.5	&	6.1	\nl
3844.909	&	328.9	&	5.6	\nl
3866.774	&	335.1	&	5.4	\nl
3867.754	&	332.1	&	4.3	\nl
3877.724	&	328.3	&	4.1	\nl
3880.810	&	339.0	&	5.0	\nl
3883.778	&	329.3	&	3.7	\nl
3888.700	&	330.7	&	4.2	\nl
3890.679	&	333.3	&	4.9	\nl
3891.682	&	333.7	&	4.6	\nl
3892.689	&	329.5	&	4.7	\nl
3893.768	&	325.2	&	4.4	\nl
3895.744	&	341.4	&	4.4	\nl
3897.749	&	332.9	&	4.6	\nl
3898.678	&	339.1	&	4.5	\nl
3901.740	&	336.0	&	4.3	\nl
3905.734	&	341.8	&	5.7	\nl
3911.730	&	333.9	&	5.8	\nl
3917.689	&	345.5	&	4.8	\nl
3938.639	&	341.2	&	16.4	\nl
3937.648	&	350.4	&	4.6	\nl
3939.631	&	338.5	&	4.7	\nl
4129.036	&	598.0	&	9.3	\nl
4131.023	&	587.5	&	9.0	\nl
4135.035	&	601.4	&	10.3	\nl
4144.998	&	619.0	&	8.0	\nl
4164.019	&	660.3	&	7.7	\nl
4176.992	&	696.4	&	7.3	\nl
4180.889	&	698.6	&	6.1	\nl
4186.887	&	711.6	&	5.5	\nl
4190.869	&	711.2	&	6.5	\nl
4191.864	&	713.7	&	5.9	\nl
4211.816	&	734.9	&	5.8	\nl
4221.789	&	744.9	&	5.7	\nl
4253.699	&	745.3	&	4.0	\nl
4282.631	&	735.8	&	4.9	\nl
4556.884	&	513.6	&	9.6	\nl
4565.914	&	510.6	&	8.7	\nl
4574.895	&	509.8	&	7.2	\nl
4580.893	&	491.0	&	7.2	\nl
4606.803	&	474.3	&	6.7	\nl
\enddata
\end{deluxetable}

\begin{deluxetable}{lccr}
\tablewidth{0in}
\tablecaption{Log of astrometric observations.\label{tab:atmlog}}
\tabletypesize{\footnotesize}
\tablehead{\colhead{Epoch}&\colhead{Date}&\colhead{N$_{obs}$}&\colhead{\textit{HST} Roll}}
\startdata
1 & 2005/Jun/15 & 4 & 58.00 \nl
2 & 2005/Jun/16 &	4 & 58.00 \nl
3 & 2005/Jun/17 &	4 & 58.00 \nl
4 & 2005/Jun/18 &	4 & 58.00 \nl
5 & 2005/Jun/19 &	4 & 58.00 \nl
6 & 2005/Jun/24 &	4 & 59.10 \nl
7 & 2006/Mar/02 &	4 & 261.00 \nl
8 & 2006/Mar/10 &	4 & 264.17 \nl
9 & 2006/Mar/15 &	4 & 266.00 \nl
10 & 2006/Mar/22 &	4 & 269.15 \nl
11 & 2006/Apr/03 &  4 & 274.00 \nl
12 & 2006/Apr/07 &	4 & 280.41 \nl
13 & 2007/Mar/03 &	4 & 261.00 \nl
14 & 2007/Mar/09 &	4 & 263.67 \nl
15 & 2007/Mar/15 &	4 & 266.00 \nl
16 & 2007/Mar/24 &	4 & 269.74 \nl
17 & 2007/Apr/01 &	4 & 274.00 \nl
18 & 2007/Apr/08 &	4 & 280.70 \nl
\enddata
\end{deluxetable}

\begin{deluxetable}{cccccc}
\tablewidth{0in}
\tablecaption{Classification spectra and photometric information for HD136118 and the astrometric reference stars. \label{tab:ast1}}
\tablehead{\colhead{Star}&\colhead{Sp~Ty }&\colhead{$V$}&\colhead{$B-V$}&\colhead{M$_{V}$}&\colhead{A$_{V}$}}
\startdata
HD\,136118	& F9V	& 6.93 &	0.55 & 3.34  & 0.0 \nl
REF-14	& K0V &	13.95 &	0.86 & 	5.88 &	0.12 \nl
REF-16 &	G0V &	12.46 &	0.73 &	4.2 &	0.45 \nl
REF-17 &	K0.5III &	13.55 &	1.13 &	0.65 &	0.21 \nl
\enddata
\end{deluxetable}

\begin{deluxetable}{ccccccccc}
\tablewidth{0in}
\tablecaption{Astrometry catalog. \label{tab:ast2}}
\tabletypesize{\scriptsize}
\rotate
\centering
\tablehead{\colhead{Star}&\colhead{\tablenotemark{a} R.A. (deg)}&  \colhead{\tablenotemark{a} Dec. (deg)} & \colhead{\tablenotemark{b}$\xi$ (arcsec)} &  \colhead{\tablenotemark{b}$\eta$ (arcsec)} & \colhead{$\mu_{\alpha}$(mas/yr)} & \colhead{$\mu_{\delta} $(mas/yr)} & \colhead{$\pi_{abs}$(mas)}  & \colhead{d (pc)}}
\startdata
HD\,136118 &229.731 & -1.592& $59.8049\pm0.0002$ & $659.3195\pm0.0002$  & $-124.06\pm0.15$ & $23.48\pm0.17$ 	& 	$19.12\pm0.23$ & $52.3\pm0.6$ \nl
REF-14 &229.739 & -1.628&$16.4201\pm0.0001$  & $783.7892\pm0.0002$ &	$ -7.39\pm0.18$ & $ 12.03\pm0.23  $	      &	 $2.75\pm0.23$&  $363\pm30$ \nl
REF-16 &229.768	 & -1.591&	$-72.6629\pm0.0001$ & $638.4385\pm0.0001$  & $1.42\pm0.13$	& $-7.21\pm0.14$      &	$2.73\pm0.18$ & $366\pm25$ \nl
REF-17 &229.723 & -1.609& $80.4884\pm0.0002$	& $721.3438\pm0.0002$  & $-7.92\pm0.28$ & $-5.72\pm0.27$      &	$0.30\pm0.02$ & $3376\pm264$ \nl
\enddata
 \tablenotetext{a}{Predicted coordinates for equinox J2000.0}   
 \tablenotetext{b}{Coordinates in the reference frame of the constrained plate (set 8, with roll = $264\degr.174$)} 
\end{deluxetable}

\begin{deluxetable}{lc}
\tablewidth{0in}
\tablecaption{HD\,136118: RV orbital parameters for a 1-companion model. \label{tab:rvparms}}
\tablehead{\colhead{Parameter} & \colhead{HD\,136118\,b}   
}
\startdata
$P$\,[days] & $1188.0\pm2.0$\nl
$T$\,[JD]&$2450614.7\pm6.3$\nl
$e$&$0.34\pm0.01$ \nl
$\omega$\,[$\degr$]&$317.1 \pm1.3$\nl
$K$\,[m/s]&$210.9\pm1.6$\nl
$M\sin{i}$\,[M$_{J}$]&$12.00\pm0.47$\nl
$\Gamma_{HET}$\,[m/s]&$485.7\pm1.8$\nl
$\Gamma_{Lick1}$\,[m/s]&$-1.1\pm3.8$ \nl
$\Gamma_{Lick2}$\,[m/s]&$-15.3\pm11.9$ \nl
\enddata
\end{deluxetable}

\begin{deluxetable}{ccccccc}
\tablewidth{0in}
\tablecaption{Statistical Quantities (SQ) for RV residuals from 1-companion orbit fit model. \label{tab:sq}}
\tablehead{
 \colhead{SQ} & \colhead{LICK} & \colhead{HET} & \colhead{ALL}
}
\startdata
Mean& -1.21 & -0.04 & -0.17\nl
Median& 0.31 & 0.73 & 0.27\nl
 $\sigma$ [m/s] & 15.89 & 8.41 & 12.03\nl
$\chi^{2}$& 21.29 & 92.42 & 117.68 \nl
DOF& 24 & 57 & 88 \nl
 $\chi^{2}_{\nu}$ & 0.89 & 1.62 & 1.34 \nl
\enddata
\end{deluxetable}

\begin{deluxetable}{ccccccc}
\tablewidth{0in}
\tablecaption{Statistical Quantities (SQ) for RV residuals from 2-companion orbit fit model. \label{tab:sq-2p}}
\tablehead{
 \colhead{SQ} & \colhead{LICK} & \colhead{HET} & \colhead{ALL}
}
\startdata
Mean& -0.61 & 0.62 & 0.34\nl
Median& -1.38 & 0.31 & 0.15\nl
 $\sigma$  [m/s] & 15.4 & 5.6 & 10.3\nl
$\chi^{2}$& 19.9 & 54.2 & 76.6 \nl
DOF& 19 & 52 & 83 \nl
 $\chi^{2}_{\nu}$& 1.05 & 1.04 & 0.92 \nl
\enddata
\end{deluxetable}

\begin{deluxetable}{lccc}
\tablewidth{0in}
\tablecaption{HD\,136118: RV Orbital Parameters for a 2-companion model. \label{tab:rvparms-2p}}
\tablehead{\colhead{Parameter} & \colhead{HD\,136118\,b} & \colhead{Nuisance Orbit}}
\startdata
$P$\,[days] &$1191\pm1.7$& $255.3\pm1.6$ \nl
$T$\,[JD] & $2450611\pm5$&$2453761\pm7$ \nl
$K$\,[m/s] & $215.8\pm1.3$& $11.2\pm1.9$ \nl
$e$& $0.353\pm0.008$& $0.50\pm0.11$ \nl
$\omega$\,[$\degr$]& $316.4\pm0.9$ &$198\pm16$ \nl
$M\sin{i}$\,[M$_{J}$]& $12.23\pm0.47$ & $0.35\pm0.07$ \nl
$a\sin{i}$\,[AU]& $2.35\pm0.16$ & $0.85\pm0.25$ \nl
$\Gamma_{HET}$\,[m/s]&\multicolumn{2}{c}{$485.5\pm1.1$} \nl
$\Gamma_{Lick1}$\,[m/s]&\multicolumn{2}{c}{$-1.4\pm3.3$} \nl
$\Gamma_{Lick2}$\,[m/s]&\multicolumn{2}{c}{$-11.4\pm9.7$} \nl
\enddata
\end{deluxetable}

\begin{deluxetable}{lc}
\tablewidth{0in}
\tablecaption{HD\,136118\,b parameters obtained from simultaneous RV and astrometry solution. \label{tab:allpars}}
\tablehead{\colhead{Parameter} & \colhead{HD\,136118\,b}   
}
\startdata
$P$\,[days] & $1190.9\pm1.2$\nl
$T$\,[JD]&$2450610.5\pm3.7$\nl
$e$&$0.352\pm0.006$ \nl
$\omega$\,[$\degr$]&$316.4 \pm0.6$\nl
$K$\,[m/s]&$215.99\pm0.92$\nl
$\alpha_{s}$\,[mas]&$1.45\pm0.25$\nl
$i$\,[$\degr$]&$163.1\pm3.0$\nl
$\Omega$\,[$\degr$]&$285\pm10$\nl
$a_{b}$\,[AU]&$2.36\pm0.05$\nl
$M_{b}$\tablenotemark{a}\,[M$_{J}$]&$42^{+11}_{-18}$\nl
$M_{b}$\tablenotemark{a}\,[M$_{\odot}$]&$0.041^{+0.010}_{-0.017}$\nl
\enddata
\tablenotetext{a}{\footnotesize It is assumed M$_{s}$ = $1.24\pm0.07$\,M$_{\odot}$}
\end{deluxetable}  
\clearpage

\begin{thebibliography}{}

\bibitem[Allende Prieto \& Lambert (1999)]{allende1999}{Allende Prieto}, C. \& {Lambert}, D.~L., 1999, \aap, 52, 555-562
\bibitem[Baraffe \etal (2001)]{baraffe2001} {Baraffe}, I. \etal, 2001, \aap, 382, 563
\bibitem[Bean \etal (2007)]{bean2007} {Bean}, J.~L. \etal, 2007, \aj, 134, 749-758
\bibitem[Benedict \etal (1998)]{benedict1998} {Benedict}, G.~F. \etal, 1998, \apj, 116, 429
\bibitem[Benedict \etal (1999)]{benedict1999} {Benedict}, G.~F. \etal, 1999, \aj, 118, 1086
\bibitem[Benedict \etal (2002a)]{benedict2002a} {Benedict}, G.~F. \etal, 2002a, \aj,  123, 473
\bibitem[Benedict \etal(2002b)]{benedict2002b} {Benedict}, G.~F. \etal, 2002b, \apj, 581, L115
\bibitem[Benedict \etal (2002c)]{benedict2002c} {Benedict}, G.~F. \etal, 2002c, \aj,124, 1695
\bibitem[Butler \etal (1996)]{butler1996} {Butler}, P. \etal, 1996, \pasp, 108, 550
\bibitem[Butler \etal (2006)]{butler2006} {Butler}, P. \etal, 2006, \apj, 646, 505
\bibitem[Chabrier \etal (2000)]{chabrier2000}{Chabrier}, G. \etal, 2000, \apj, 542, 464
\bibitem[Chambers (1999)]{chambers1999} {Chambers}, J.~E., 1999, \mnras, 304, 793-799
\bibitem[Fischer \etal (2002)]{fischer2002} {Fischer}, D.~A. \etal, 2002, \pasp, 114, 529-535
\bibitem[Gonzalez \& Laws (2007)]{gonzalez2007} {Gonzalez}, G. and {Laws}, C., 2007, \mnras, 378, 1141
\bibitem[Gregory (2007)]{gregory2007} {Gregory}, P.~C., 2007, \mnras, 381, 1607-1616
\bibitem[Grether \& Lineweaver (2006)]{grether2006} {Grether}, D. \& {Lineweaver}, C.~H., 2006, \aj, 640, 1051-1062
\bibitem[Henry \etal (2000)]{henry2000} {Henry}, G.~W. \etal, 2000, \apj, 529, L41-L44
\bibitem[Jefferys, Fitzpatrick \& McArthur (1988)]{jefferys1988} {Jefferys}, W.~H., {Fitzpatrick}, M.~J., \& {McArthur}, B.~E., 1988, Celestial Mechanics, 41, 39
\bibitem[McArthur \etal (1997)]{mcarthur1997}{McArthur}, B. ~E., {Benedict}, G.~F., {Jefferys}, W.,~H. and {Nelan}, E., 1997, in The 1997 HST Calibration Workshop, ed. S. Casertano, R. Jedrzejewski, C. D. Keyes, \& M. Stevens (Baltimore : STScI), 472
\bibitem[McArthur \etal (2001)]{mcarthur2001} {McArthur}, B.~E. \etal, 2001, \apj, 560, 907
\bibitem[Nelan \etal (2007)]{nelan2007} {Nelan}, E. \etal, 2007, Fine Guidance Sensor Handbook, Ver. 16.0 (Baltimore: STSCcI)
\bibitem[Perryman (1997)]{perryman1997} {Perryman}, M.~A.~C., 1997, \aap, 323, L49-L52
\bibitem[Pourbaix \& Jorrisen (2000)]{pourbaix2000} {Pourbaix}, D. \& {Jorrisen}, A., 2000, \aap, 145,161
\bibitem[Reffert \& Quirrenbach (2006)]{reffert2006} {Reffert}, S. \& {Quirrenbach}, A., 2006, \aap, 449,699-702
\bibitem[Saar \& Donahue (1997)]{saar1997}{Saar}, S.~E. \& {Donahue}, R.~A., 1997, \apj, 485, 319
\bibitem[Saffe \etal (2005)]{saffe2005} {Saffe}, C. \etal, 2005, Revista Mexicana de Astronomia y Astrofisica, 41, 415-421
\bibitem[Scargle (1982)]{scargle1982} {Scargle}, J.~D., 1982, \apj, 263, 835
\bibitem[HRS; Tull (1998)]{tull1998} {Tull}, R.~G., 1998, Proc. SPIE, 3355, 387
\bibitem[Van Leeuwen (2007)]{vanleeuwen2007} {van Leeuwen}, F., 2007, \aap, 474,653-664
\bibitem[Whipple \etal (1995)]{whipple1995}{Whipple}, A. ~L. \etal, 1995, in Calibrating Hubble Space Telescope : Post-Servicing Mission, ed. A. Koratkar \& C. Leitherer (Baltimore : STScI),119
\bibitem[Zacharias \etal (2004)]{zacharias2004} {Zacharias}, N. \etal, 2004, \aj, 127, 3043-3059

\end{thebibliography}
\end{document}